\pdfoutput=1

\documentclass[twocolumn]{aastex63}

\usepackage{amsmath}
\usepackage{epsfig}
\usepackage{amssymb}
\usepackage{graphicx}
\usepackage{color}
\usepackage{natbib}
\usepackage{ulem}
\usepackage[usestackEOL]{stackengine}

\def\gtaprx {\lower .1ex\hbox{\rlap{\raise .6ex\hbox{\hskip .3ex
	{\ifmmode{\scriptscriptstyle >}\else
		{$\scriptscriptstyle >$}\fi}}}
	\kern -.4ex{\ifmmode{\scriptscriptstyle \sim}\else
		{$\scriptscriptstyle\sim$}\fi}}}
\def\ltaprx {\lower .1ex\hbox{\rlap{\raise .6ex\hbox{\hskip .3ex
	{\ifmmode{\scriptscriptstyle <}\else
		{$\scriptscriptstyle <$}\fi}}}
	\kern -.4ex{\ifmmode{\scriptscriptstyle \sim}\else
		{$\scriptscriptstyle\sim$}\fi}}}

\newcommand{\cutt}[1]{\textcolor{blue}{}}

\newcommand{\Ms}{{\ensuremath{\mathrm{M}_{\odot} }}}
\newcommand{\Ls}{{\ensuremath{L_{\odot} }}}

\newcommand{\C}{{\ensuremath{^{12}\mathrm{C}}}}

\newcommand{\N}{{\ensuremath{^{14} \mathrm{N}}} }

\submitjournal{ApJ}

\shorttitle{Supermassive Stars in Cosmological Flows}
\shortauthors{Woods et al.}

\graphicspath{{./}{figures/}}

\begin{document}

\title{On the Evolution of Supermassive Primordial Stars in Cosmological Flows}

\correspondingauthor{Tyrone E. Woods}
\email{tyrone.woods@nrc-cnrc.gc.ca}

\author[0000-0003-1428-5775]{Tyrone E. Woods}
\affiliation{National Research Council of Canada, Herzberg Astronomy \& Astrophysics Research Centre, 5071 West Saanich Road, Victoria, BC V9E 2E7, Canada}
\affiliation{Monash Centre for Astrophysics, School of Physics and Astronomy, Monash University, VIC 3800, Australia}

\author[0000-0002-5293-699X]{Samuel Patrick}
\affiliation{Institute of Cosmology and Gravitation, University of Portsmouth, Portsmouth PO1 3FX, UK}

\author[0000-0002-6139-2226]{Jacob S. Elford}
\affiliation{Institute of Cosmology and Gravitation, University of Portsmouth, Portsmouth PO1 3FX, UK}

\author[0000-0002-1463-267X]{Daniel J. Whalen}
\affiliation{Institute of Cosmology and Gravitation, University of Portsmouth, Portsmouth PO1 3FX, UK}
\affiliation{Ida Pfeiffer Professor, University of Vienna, Department of Astrophysics, Tuerkenschanzstrasse 17, 1180, Vienna, Austria}

\author[0000-0002-3684-1325]{Alexander Heger}
\affiliation{Monash Centre for Astrophysics, School of Physics and Astronomy, Monash University, VIC 3800, Australia}
\affiliation{ARC Centre of Excellence for Gravitational Wave Discovery (OzGrav), Melbourne, Australia}
\affiliation{ARC Centre of Excellence for Astrophysics in Three Dimensions (ASTRO-3D), Australia}
\affiliation{Joint Institute for Nuclear Astrophysics, 1 Cyclotron Laboratory, National Superconducting Cyclotron Laboratory, Michigan State University, East Lansing, MI 48824-1321, USA}

\begin{abstract}

Primordial supermassive stars (SMSs) formed in atomic-cooling halos at $z \sim$ 15 - 20 are leading candidates for the seeds of the first quasars. Past numerical studies of the evolution of SMSs have typically assumed constant accretion rates rather than the highly variable flows in which they form.
We model the evolution of SMSs in the cosmological flows that create them using the {\sc Kepler} stellar evolution and implicit hydrodynamics code.  We find that they reach masses of $1 - 2 \times 10^5$ \Ms\ before undergoing direct-collapse to black holes (DCBHs) during or at the end of their main-sequence hydrogen burning, at 1 - 1.5 Myr, regardless of halo mass, spin, or merger history. We also find that realistic, highly-variable accretion histories allow for a much greater diversity of supermassive stellar structures, including in some cases largely thermally relaxed objects, which may provide a significant source of radiative feedback. Our models indicate that the accretion histories predicted for purely atomic-cooling halos may impose a narrow spectrum of masses on the seeds of the first massive quasars, however further studies incorporating realistic feedback will be essential in order to confirm whether or not this holds true in all cases.  Our results also indicate that multiple SMSs at disparate stages of evolution can form in these halos, raising the possibility of SMS binaries and supermassive X-ray binaries (SMXBs), as well as DCBH mergers which could be detected by LISA.

\end{abstract}


\section{Introduction}

Supermassive stars (SMS), as the progenitors of direct collapse black holes, are now one of the leading candidates for the seeds of the first massive quasars in the universe. Over 200 such objects have now been confirmed at $z >$ 6 \citep{ivh20,quasardataset}, with 8 known quasars at $z>7$ (e.g., \citealt{mort11,wu15,ban18,smidt18,mats19,wang21}).  In the canonical formation scenario, a primordial halo grows to masses of 10$^7$ - 10$^8$ \Ms\ without ever having formed stars, either because it is immersed in a strong Lyman-Werner background from nearby Population III (Pop III) stars \citep[e.g.,][]{latif15a,agarw15}, strong dynamical heating resulting from a rapid halo assembly history \citep{yahs03, Fernandez2014, wise19}, and/or due to supersonic baryon streaming flows \citep{Tanaka2014,hir17,srg17}.  At these masses, the virial temperature of the halo reaches $\sim$ 10$^4$ K and triggers rapid atomic H cooling, which causes gas to collapse at rates of up to $\sim$0.1--1 \Ms\ yr$^{-1}$ \citep{bl03,wta08, sbh10, bec15}.  These flows flatten into an accretion disk that builds up at least one SMS at its center \citep[e.g.][]{ln06,rh09b,latif13a,rd18}, although recent simulations suggest binaries or even small multiples of SMSs are possible (\citealt{rjh14, latif20a, pat20a} -- for recent reviews, see \citealt{rosa17,titans,sb19,ivh20,hle20}). 

Stellar evolution models show that these supermassive objects do not immediately collapse, but survive as core hydrogen-burning stars that can reach masses of several 10$^5$ \Ms\ when accreting rapidly before dying as direct-collapse black holes (DCBHs) via the general relativistic (GR) instability (e.g., \citealt{ful86} and references therein; \citealt{um16,tyr17,hle18a,hle18b,herr21a}). 
Model calculations suggest that in some rare circumstances SMSs may explode at the onset of core hydrogen burning in the case of monolithic collapse \citep[see discussion in][]{tyr20a} or during core helium burning \citep{jet13a,chen14b}, although see \citet{nag20}. Most of the SMS evolution models find that they evolve as cool, red supergiants along the Hayashi track because of H$^{-}$ opacity in their atmospheres \citep[e.g.,][]{hos12,hos13}, so they are not thought to be bright sources of ionizing UV radiation that could alter accretion onto the star \citep{ard18,luo18}, although Ly$\alpha$ radiation pressure might affect flows on small scales \citep{aaron17}.  Pop III SMS, nevertheless, have extremely large luminosities that could be detected in the near infrared (NIR) by the {\em James Webb Space Telescope} ({\em JWST}) and large ground-based telescopes in the coming decade \citep{hos13,sur18a,sur19a,wet20a}.

Most studies of Pop III SMS to date, however have only considered constant accretion rates, not the rapidly varying flows in which the stars actually evolve.  \citet{sak15} examined the evolution of Pop III SMS in idealized, bursty accretion that approximated dense clumps in the disk due to fragmentation, and found that the star could exhibit temporary blue, hot phases before settling down onto cooler, redder evolutionary tracks.  These hotter episodes occurred when quiescent times between bursts were longer than $\sim$ 1,000 yr and the accretion rates between bursts were low because they gave the star's outer layers time to thermally relax into more compact states.  \citet{sak15} followed up this study with accretion rates taken from idealized two-dimensional (2D) primordial disk simulations.  They found that in practice accreting protostars will remain cool and red allowing them to reach the supermassive regime \citep{sak16b}.

New cosmological simulations have followed the evolution of atomically-cooling halos over the times required for an SMS to form and then collapse to a DCBH \citep[e.g.,][]{pat20a}.  They show that the accretion disk of the star can be highly turbulent and is prone to episodes of fragmentation, in which large clumps frequently collide with the star.  These accretion rates are quite different than those considered in any SMS model so far \citep[see also][]{bec15,bec18}.  How Pop III SMS evolve in the turbulent, clumpy flows that create them is crucial to understanding their temperatures, luminosities, actual growth rates and masses at collapse.  Intermittent, high surface temperatures could reduce accretion rates onto the star, prolonging its life and reducing its final mass.  These processes in turn determine the prospects for the detection of the star and the BH it will ultimately form, and therefore the first quasars at birth.  Semi-analytic models indicate any massive clumps formed from fragmentation in the disk will coalesce and merge before entering the main sequence as ionizing sources \citep{ih14}; however, if turbulence and quiescent accretion lead to extended periods in which the SMS itself is a strong source of ionizing UV photons, then radiation hydrodynamical models that are fully coupled to stellar evolution models may be required to capture the true properties of the star and its mass at death.  In the present work, we have evolved Pop III SMS in their natal cosmological flows.  These flows were taken from a sample of simulated atomic-cooling halos that span the likely range of assembly histories \citep{pat20a}.  In Section 2, we discuss our cosmological models, their accretion rates, and our stellar evolution models.

In Section 3, we examine SMS evolution in these flows and analyze their properties.  As new cosmological simulations have revealed the possibility of supermassive binaries, we conclude in Section 4 with a brief discussion of the prospects for interactions between SMSs in the early Universe.
\vskip0.5cm

\section{Numerical Method}

We draw accretion rates for the first disk to form at the centers of the eight halos examined in the {\sc enzo} adaptive mesh refinement (AMR) cosmology code as given by \cite{pat20a}.  These rates were then used to evolve Pop III SMS in the {\sc Kepler} stellar evolution code, as described below.

\subsection{Enzo Simulations}

The eight halos for which we tallied central accretion rates were simulated in 1.5 $h^{-1}$ Mpc boxes in {\sc enzo} \citep{enzo}, which self-consistently evolves gas, dark matter, and nonequilibrium primordial gas chemistry and cooling. Each run had a $(256)^3$ root grid with three nested grids centered on the halo and up to 15 levels of refinement for a maximum resolution of 0.014 pc.  To simulate the classic SMS formation scenario of isothermal collapse in high Lyman-Werner UV backgrounds, these simulations were limited to six primordial gas species (H, H$^+$, He, He$^+$, He$^{++}$, and e$^-$) with no H$_2$ chemistry \citep[though see][for recent works investigating the cases of only a very modest Lyman-Werner background and with local UV feedback from the SMS itself, respectively]{Regan20, sak20}.  Cooling due to collisional excitation and ionization of H and He, bremsstrahlung, and inverse Compton cooling by the cosmic microwave background were all included.

\begin{figure*}
    \centering
    \includegraphics[width=0.9\textwidth]{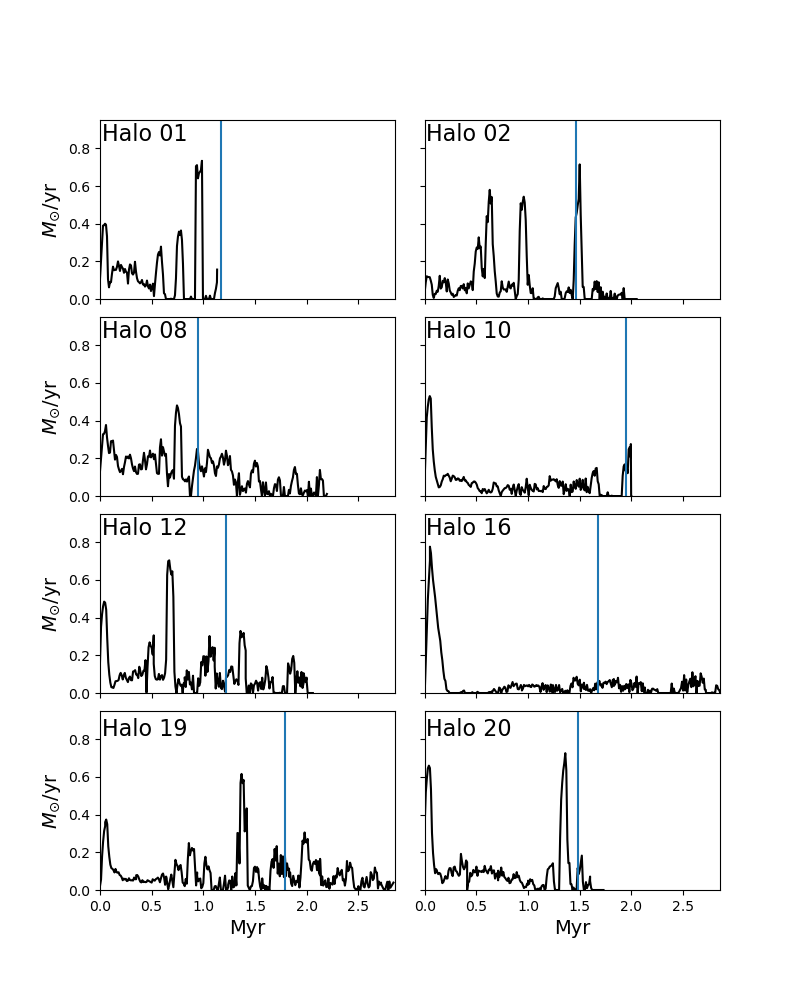}
    \caption{SMS accretion rates from \citet{pat20a}.  The blue vertical lines indicate the times when the stars collapse in our {\sc Kepler} runs (see Section 3).  Only the main disk in each halo is considered here.
    \label{fig:rates}}
\end{figure*}

Each halo was evolved for 2 - 3 Myr after the onset of collapse due to atomic cooling, which occurred at redshifts $z =$ 13.9 - 20.4 at masses of $1.2 \times 10^7 ~\Ms$ - $8.5 \times 10^7$ \Ms.  These halos were chosen to bracket their likely range of spin parameters, $\lambda$, where 
\begin{equation}
\lambda \, = \, \frac{|J|}{\sqrt{2GMR}},
\end{equation}
and $J$, $R$, and $M$ are the angular momentum, virial radius and virial mass of the halo, respectively.  They were also chosen to include a range of assembly histories ranging from growth primarily by accretion to growth by major mergers. We refer the reader to \cite{pat20a} for further details regarding the halo selection criteria. Our ensemble thus yields SMS accretion rates for a variety of primordial environments.  We extracted these rates at 10 kyr intervals from the onset of disk formation for all eight halos. Accretion rates onto the central objects were found by summing the total mass flux through a sphere 0.134pc in radius (resolved with at least 20 zones), centred within the disk. At times, the accretion rate as modelled in {\sc enzo} can become negative if turbulence drives gas out of the center of the disk. When this happens, we assume that accretion onto any central object formed is halted, i.e., the innermost accretion disk is emptied. We set the accretion rate to zero for our {\sc Kepler} simulation during such intervals. The resulting input accretion histories are plotted in Figure~\ref{fig:rates}. It is important to clarify here that fragmentation on smaller scales unresolved by our simulation could in principle produce smaller stars and lower mean accretion rates, as found in e.g., \citet{Regan20}; we may note, however, that in contrast to either halo explored in their work, the halo simulations in this study are all implicitly assumed to be embedded in a very strong Lyman-Werner background, and that disk fragmentation in such an environment on scales $\lesssim \rm{few} \times 10^{-2}$ pc is expected to produce fragments which will merge with the central object before reaching the zero-age main sequence \citep{ih14}.

\begin{figure} 
\begin{center}
\begin{tabular}{cc}
\epsfig{file=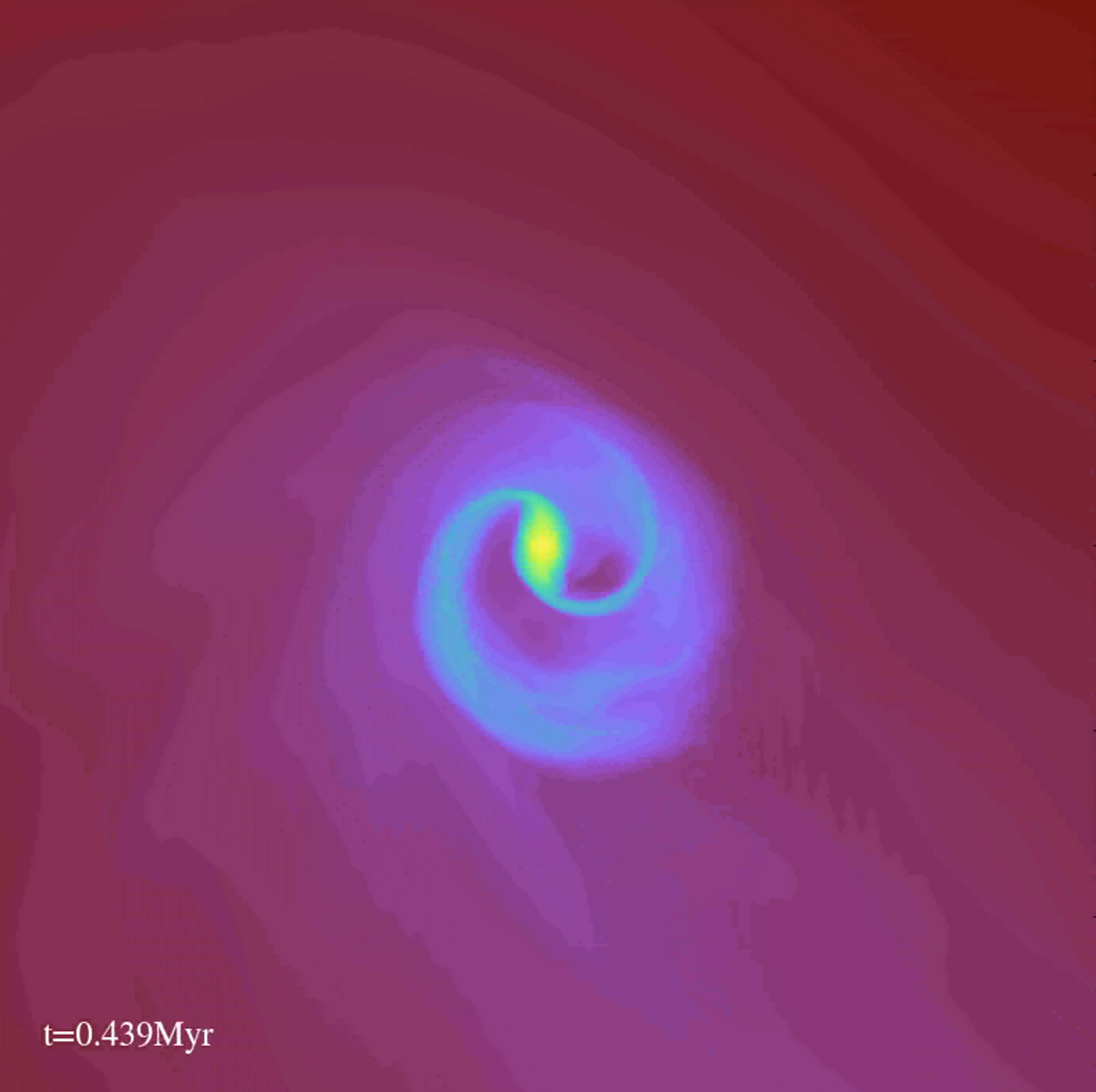,width=0.45\linewidth,clip=}  &
\epsfig{file=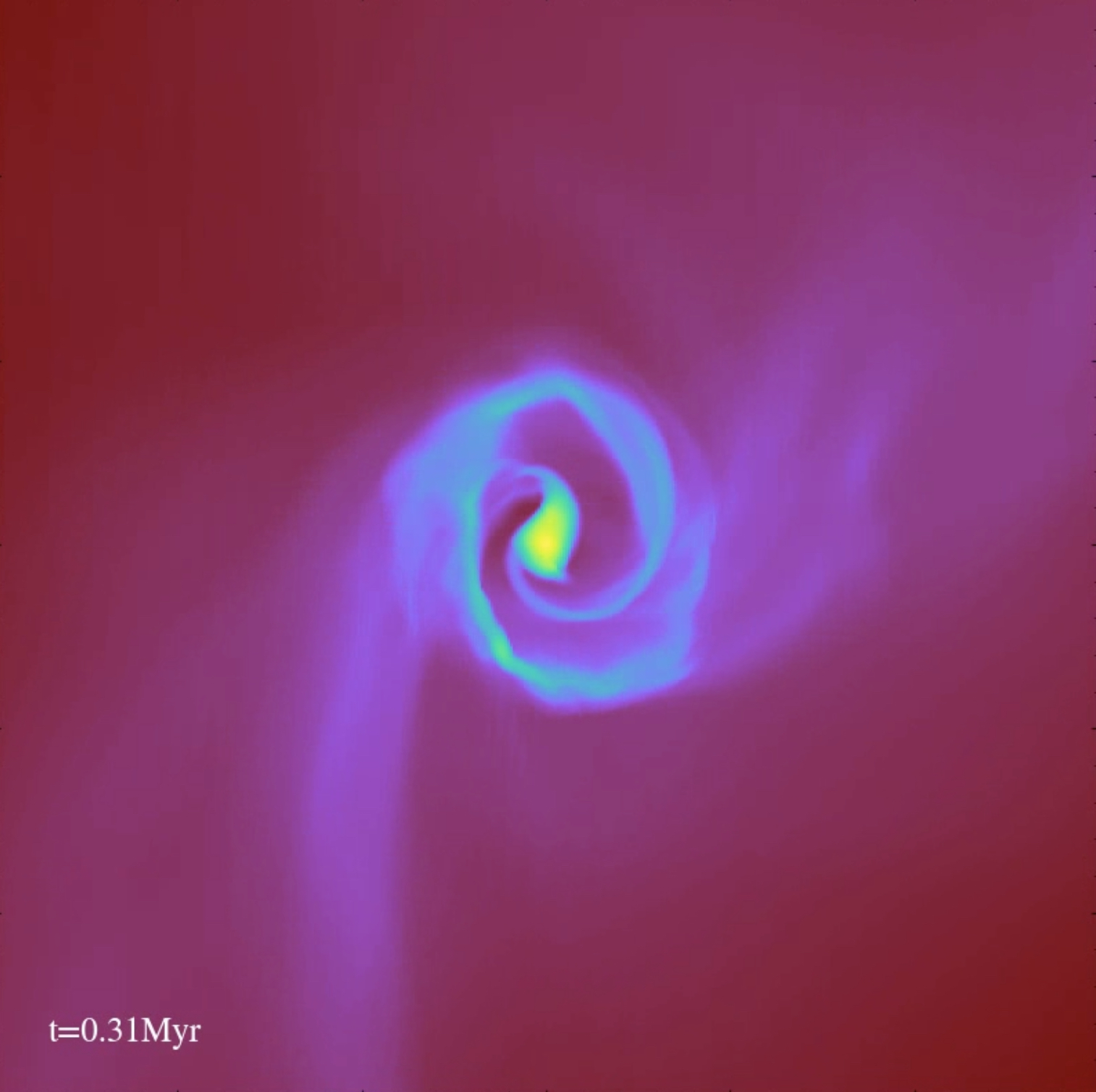,width=0.45\linewidth,clip=}  \\
\epsfig{file=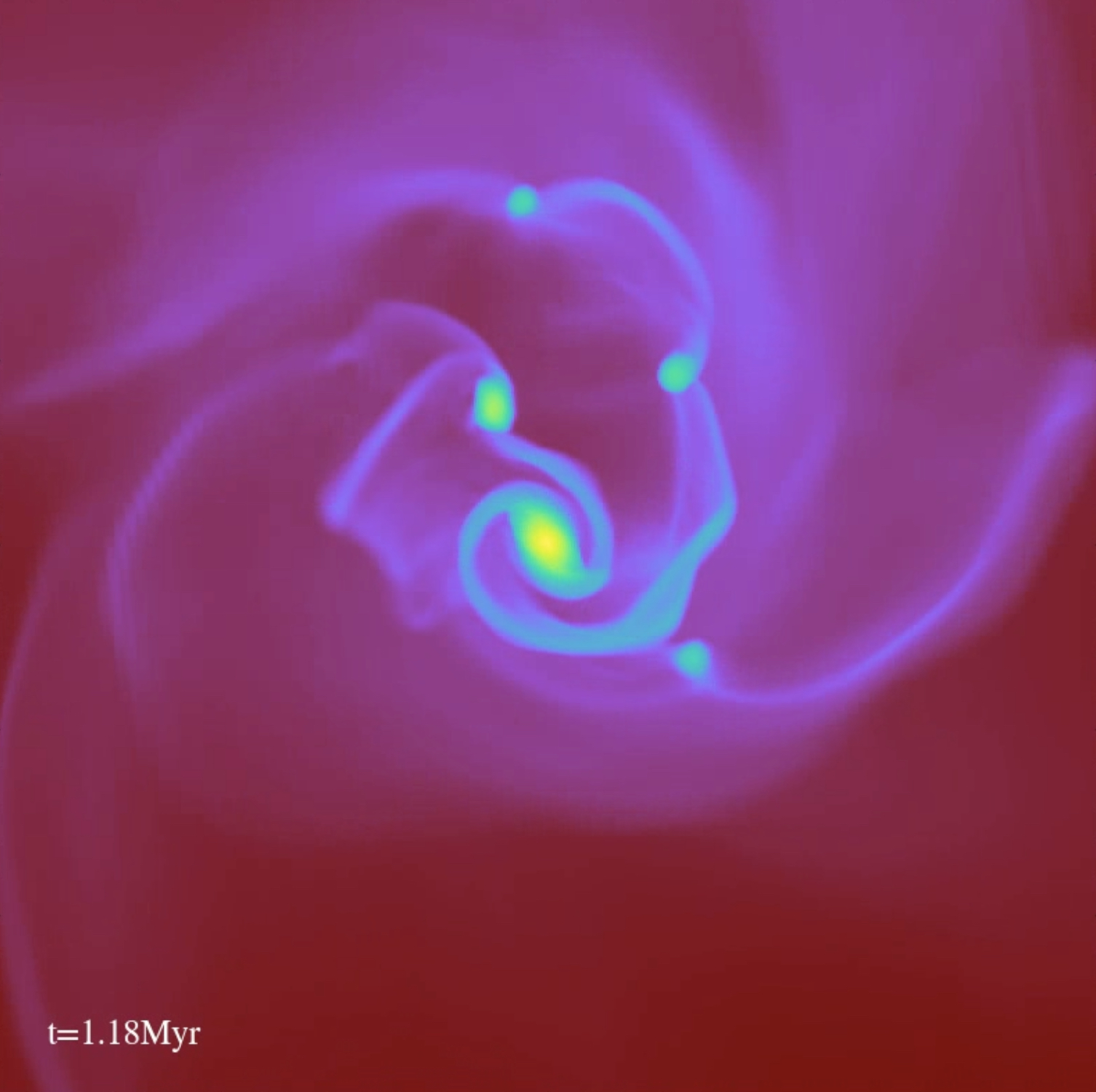,width=0.45\linewidth,clip=}  &
\epsfig{file=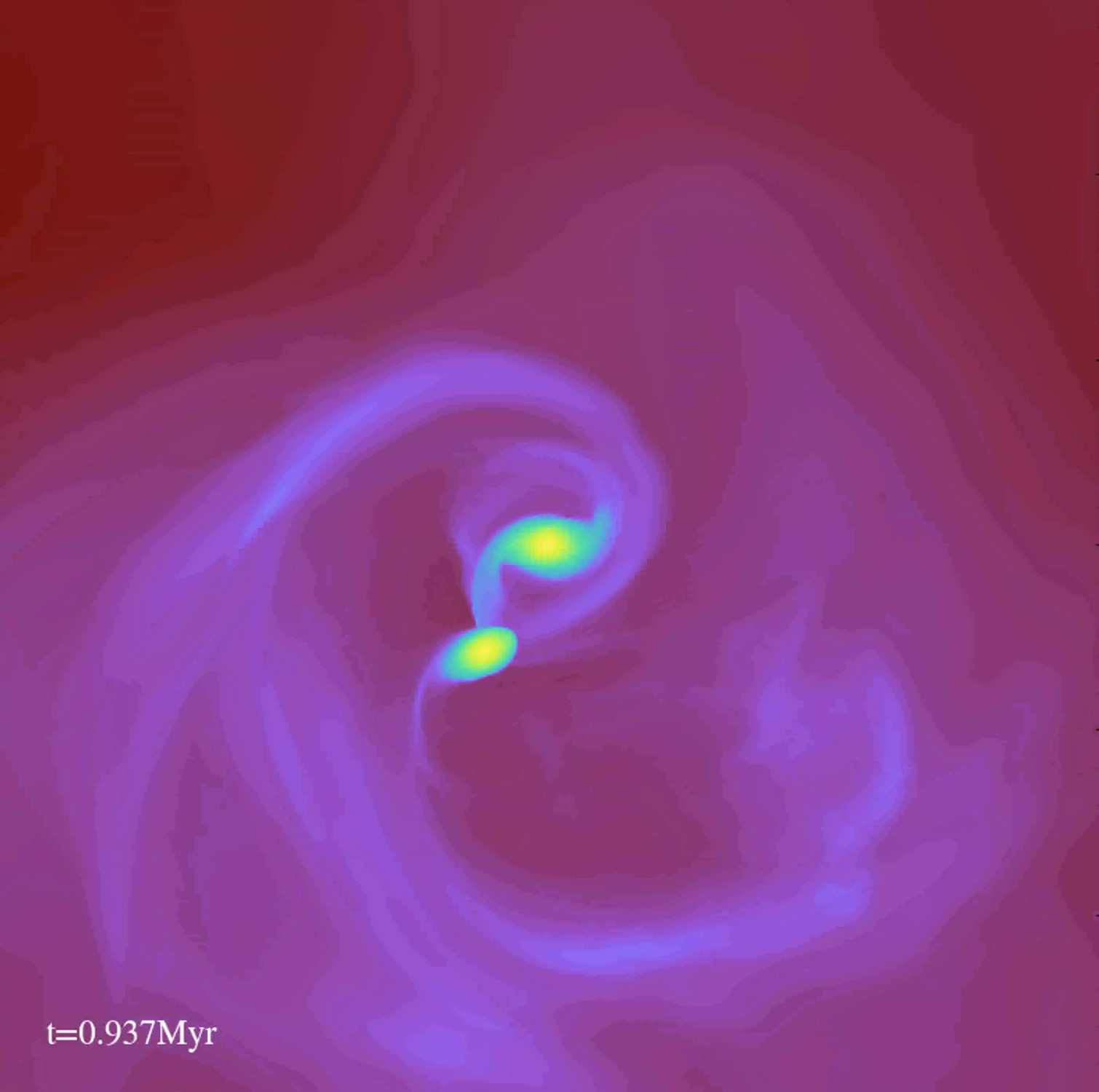,width=0.45\linewidth,clip=}  
\end{tabular}
\end{center}
\caption{Density projections of accretion disks in various states of flow in the atomically-cooled halos in \citet{pat20a}.  Top left:  relatively smooth accretion in \texttt{Halo 10} at 0.439 Myr.  Top right:  turbulent accretion flows in \texttt{Halo 20} at 0.31 Myr.  Bottom left:  fragmentation of the disk in \texttt{Halo 12} at 1.18 Myr.  Bottom right:  the onset of the collision of a satellite disk with the main disk in \texttt{Halo 01} at 0.937 Myr.  Each image is 6 pc on a side. \label{fig:disks}}
\end{figure}

Accretion begins with an initial surge, ranging from 0.3 - 1 \Ms\ yr$^{-1}$, that lasts for 200 - 300 kyr.  The surge coincides with the formation of the disk, which has an initial diameter of $\sim$ 0.5 pc.  Accretion thereafter can then be relatively smooth (as in \texttt{Halo 10} in the upper left panel of Figure~\ref{fig:disks}), highly turbulent (as in \texttt{Halo 20} 
in the upper right panel of Figure~\ref{fig:disks}) or clumpy because of the fragmentation of the disk (as in \texttt{Halo 12} in the lower left panel of Figure~\ref{fig:disks}).  The large jumps in accretion rate at intermediate times are generally due to the fragmentation of the disk and subsequent collision of the clumps with its center, as at 0.7 Myr in \texttt{Halo 02}.  The large peak at 1.0 Myr in \texttt{Halo 01} (and at 1.4 Myr in \texttt{Halo 19}) is due to the collision of a satellite disk that also hosts an SMS with the main disk, as shown in the lower right panel of Figure~\ref{fig:disks}.

\subsection{{\sc Kepler} Models}

We evolve each star in the one-dimensional Lagrangian stellar evolution and hydrodynamics code {\sc Kepler} \citep{kep1,kep2} using the post-Newtonian approximation implemented by \citet{ful86}.  Specifically, {\sc Kepler} solves the angular momentum and energy equations with first-order (post-Newtonian) GR corrections to gravity:

\begin{eqnarray}\label{cons_mom}
\frac{dv}{dt} & = & 4\pi r^{2} \frac{\partial P}{\partial m_{r}} + \frac{4\pi}{r}\frac{\partial Q}{\partial m_{r}} - \frac{G_{\rm{rel}}m_{r}}{r^{2}} \\
\frac{du}{dt} & = & -4\pi P \frac{\partial}{\partial m_{r}} (vr^{2}) + 4\pi Q \frac{\partial}{\partial m_{r}}\left(\frac{v}{r}\right) - \frac{\partial L}{\partial m_{r}} + \epsilon ~~~
\end{eqnarray}

\noindent The terms on the right-hand side of Equation~\ref{cons_mom} are the acceleration due to pressure gradients, viscous drag and gravity, respectively, while those on the right-hand side of Equation~3 are the energy flux due to work, viscous dissipation, radiative flux, and nuclear burning.  The post-Newtonian correction to gravity is implemented by modifying the gravitational constant:

\begin{equation}
G_{\rm{rel}} = G \left(1 + \frac{P}{\rho c^{2}} + \frac{4\pi Pr^{3}}{m_{r}c^{2}}\right)\left(1 - \frac{2Gm_{r}}{rc^{2}}\right)^{-1}
\end{equation}

\noindent The factor $Q$ in the viscous term is

\begin{equation}
Q = \frac{4}{3}\,\eta_{\nu}r^4\frac{\partial}{\partial r}\left(\frac{v}{r}\right)\;,
\end{equation}

\noindent where $\eta _{\nu}$ is the dynamic viscosity from \citet{kep1}, defined as:
\begin{equation}
    \eta _{\nu} = \eta _{R} + \frac{3}{4}l_{1}\rho c_{s} + \frac{3}{4}l_{2}^{2} \rm{max}(0, -\nabla \cdot \bf{v})
\end{equation}

\noindent which includes the real ($\eta _{R}$) and artificial viscosity, with the latter used to dampen acoustic oscillations during quiescent phases of the evolution of a star. Here we choose the canonical values $l_{1} = 0.1\Delta r$ and $l_{2} = 2\Delta r$ for Kepler simulations, where $\Delta r$ is the local width of a zone.

Nuclear-burning is evolved with an adaptive network that is implicitly coupled to the hydrodynamics \citep{kep3}.  We use a Helmholtz-like equation of state that includes electron-positron pair production, relativistic and non-relativistic degenerate and non-degenerate electrons, and radiation \citep{ts00}.  Time-dependent convection as described in \cite{kep1} is included in our runs, as is convective heat transport when a zone satisfies the Ledoux criterion.  

We initialize our models as 10 \Ms, $n =$ 3 polytropes with primordial compositions, and with central densities $\rho_{\mathrm{c}} =$ 10$^{-3}$ g cm$^{-3}$ and temperatures $T_{\mathrm{c}} =$ $1.2 \times 10^6$ K, which are capable of sustaining deuterium burning. Varying the initial mass and entropy significantly does not have a substantive impact
on the subsequent evolution of the SMS, so long
as it is small relative to the final SMS mass, given
that the core quickly becomes convective; see discussion in \citet{tyr17} and \citet{hle18b}.  The accretion flow onto the star is also primordial in composition.  We match its entropy to that of the surface of the star, which neglects the luminosity due to the accretion shock, but at 1 \Ms\ yr$^{-1}$ this luminosity is on the order of 10$^4$ \Ls, which is negligible in comparison to that of the star.  Studies that include this luminosity at the base of the photosphere of the SMS find little change in its evolution above a few thousand \Ms\ \citep[e.g.,][]{hos13}. We also do not include the accretion of angular momentum, restricting our attention to non-rotating models \citep[though see][for more on rotating SMSs]{hle18a,hle19}. Additional details on our treatment of accretion can be found in \citet{kep3} and \citet{tyr17}.

{\sc Kepler} partitions each star into a large number of zones (here, up to $\sim$8,000) with especially high mass resolution near its center ($\sim$ 10$^{30}$ g/zone) and outer layers ($\sim$ 10$^{26}$ g/zone).  In each model, the code makes an arbitrarily large initial guess for each time step, and then iterates to identify the largest step the code can take without exceeding preset limits on the change in fractional radius, temperature, luminosity, or density anywhere on the grid, and while still following the emergence of shocks.  This approach enables us to follow the long-term evolution of the star over thermal and nuclear timescales, dropping to short time steps to resolve hydrodynamic timescales only when instabilities appear. This includes the onset of the post-Newtonian or GR instability \citep{chandra64}.  Recently, \citet{Haemmerle20} confirmed that {\sc Kepler} predicts collapse at masses that are consistent with those expected from careful analytic estimates. We ignore mass loss due to winds, which are thought to be negligible in primordial stars \citep{vink01} and in any event could not overcome the ram pressure of infall.  We also do not consider pulsations and pulsational mass losses here \citep{bhw01}.

\section{Results}

\begin{figure*} 
\begin{center}
\begin{tabular}{cc}
\stackinset{c}{}{t}{0cm}{\textbf{Halo 01}}{\epsfig{file=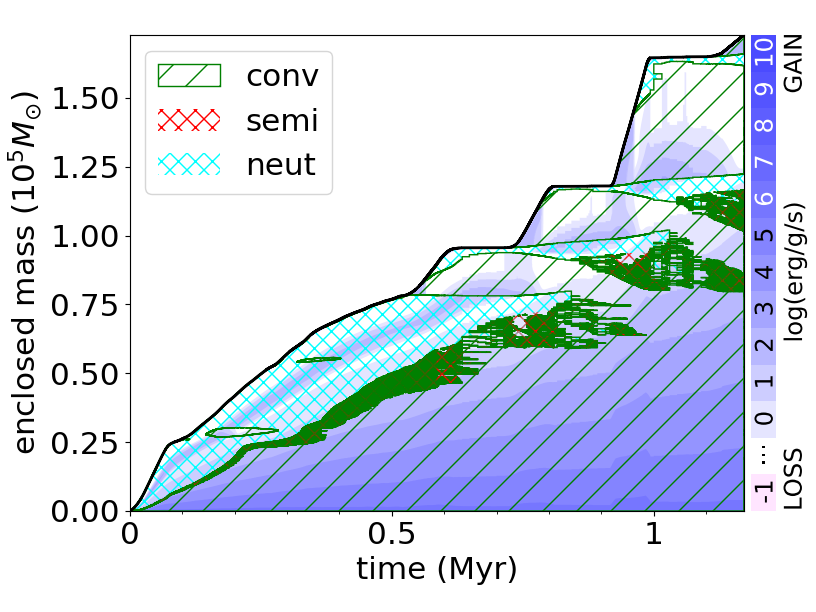,width=0.4\linewidth,clip=}} &
\stackinset{c}{}{t}{0cm}{\textbf{Halo 02}}{\epsfig{file=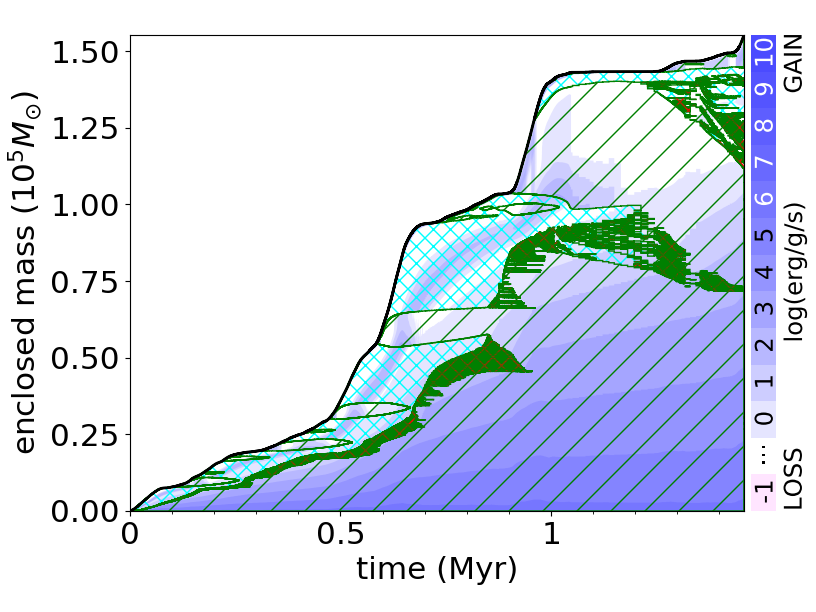,width=0.4\linewidth,clip=}}  \\
\stackinset{c}{}{t}{0cm}{\textbf{Halo 08}}{\epsfig{file=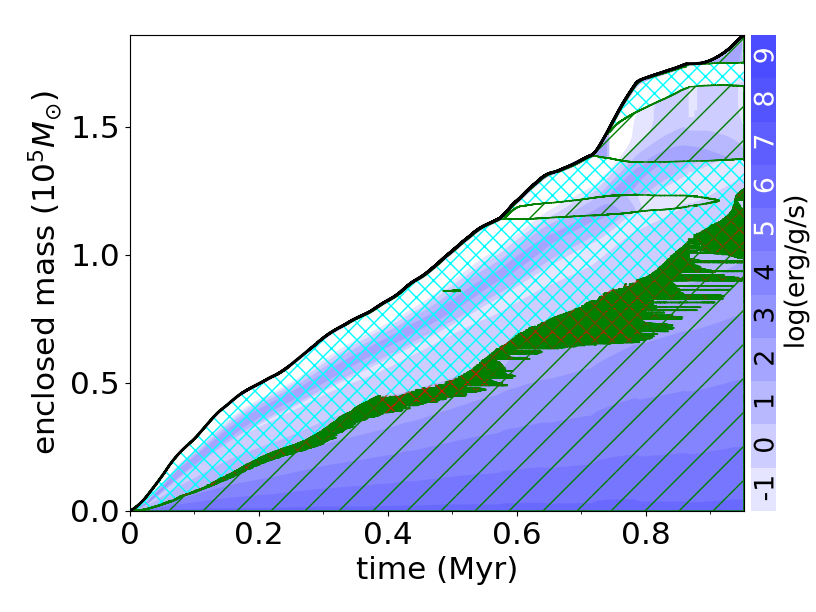,width=0.4\linewidth,clip=}}  &
\stackinset{c}{}{t}{0cm}{\textbf{Halo 10}}{\epsfig{file=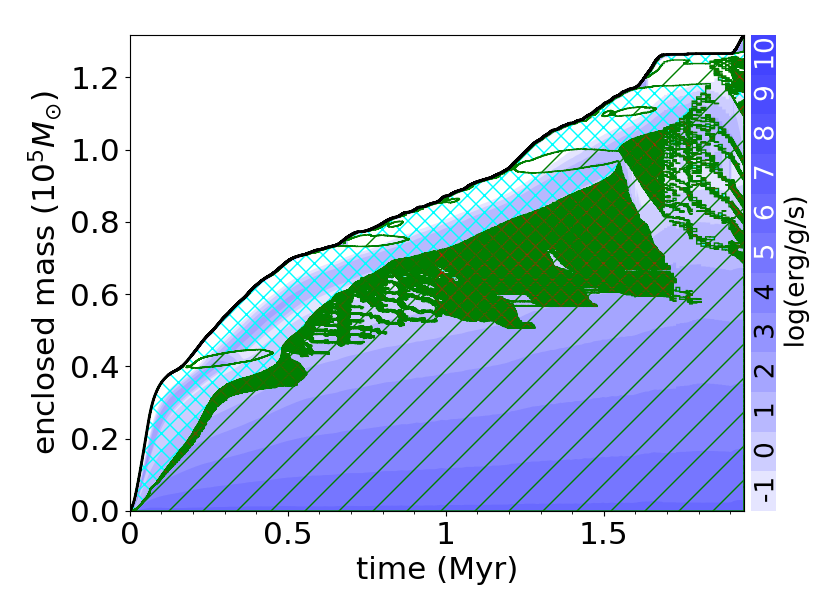,width=0.4\linewidth,clip=}}  \\
\stackinset{c}{}{t}{0cm}{\textbf{Halo 12}}{\epsfig{file=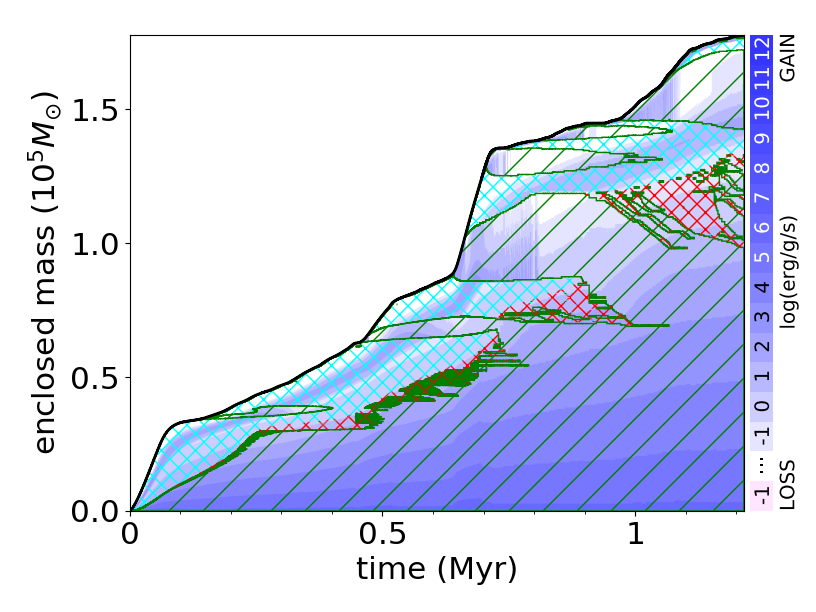,width=0.4\linewidth,clip=}}  &
\stackinset{c}{}{t}{0cm}{\textbf{Halo 16}}{\epsfig{file=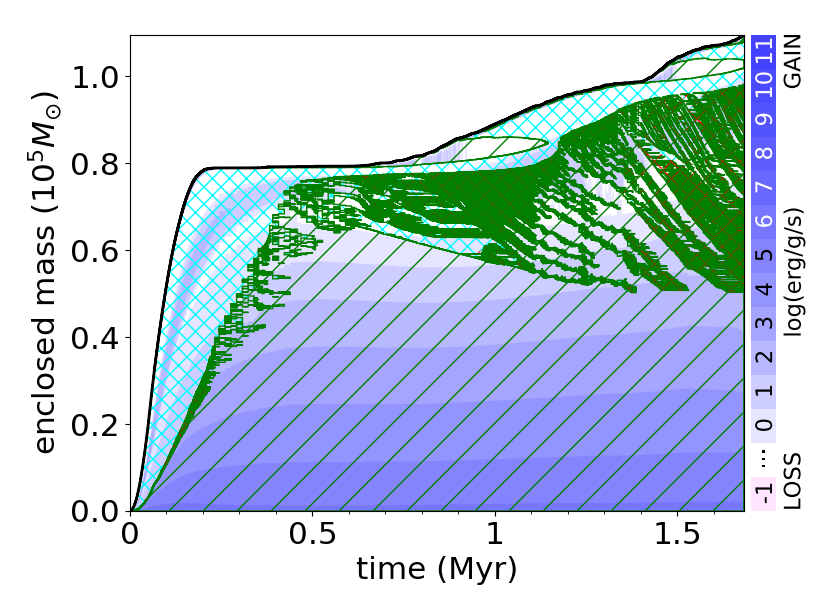,width=0.4\linewidth,clip=}}  \\
\stackinset{c}{}{t}{0cm}{\textbf{Halo 19}}{\epsfig{file=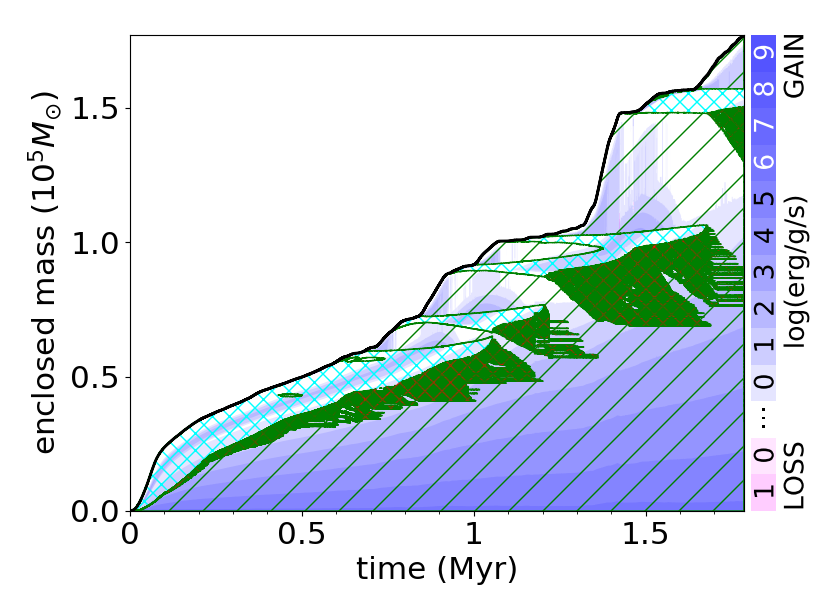,width=0.4\linewidth,clip=}}  &
\stackinset{c}{}{t}{0cm}{\textbf{Halo 20}}{\epsfig{file=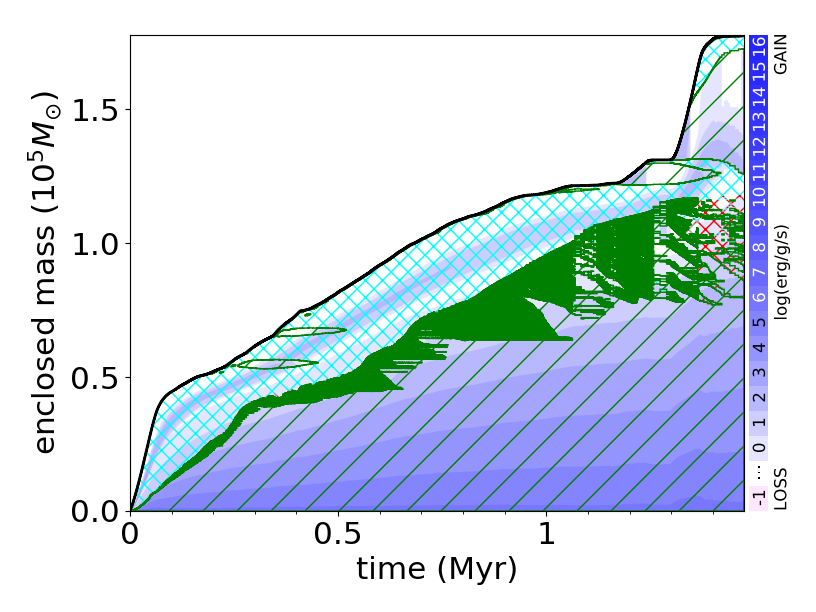,width=0.4\linewidth,clip=}}  \\
\end{tabular}
\end{center}
\vskip-0.5cm
\caption{Kippenhahn diagrams illustrating energy generation, energy transport, and stellar structure as a function of time and mass coordinate for the stars in our study. Figure captions: ``conv'' (green single-line hashing) denotes convective regions, ``semi'' (red 'x' cross-hashing) denotes semi-convective regions, and ``neut'' (light blue 'x' cross-hashing) denotes neutral regions. 
The specific energy generation rate (blue shading) at each mass coordinate in the star is indicated by the color axis \citep[see, e.g., the description in][]{tyr17}.  \label{Kippenhahn}
}
\end{figure*}

The evolution of the star in the main disk in each halo, along with its total mass and energy generation and transport within the star, are shown in the Kippenhahn diagrams in Figure~\ref{Kippenhahn}.  We discuss the salient features of each model in Section 3.1 and infer general aspects of SMS formation and their connection to the properties of their host halos in Section 3.2.

\subsection{Individual Halos}

In \texttt{Halo 01}, the initial surge in accretion rate due to the formation of the disk peaks at $\sim$ 0.4 \Ms\ yr$^{-1}$ in the first 100 kyr, before subsiding somewhat. 
Consequently, the SMS initially develops a large, high-entropy radiative envelope surrounding its convective, nuclear-burning core, much as it would in the constant accretion rate case \citep[e.g.,][]{hos13}. This outer radiative envelope is only very marginally stable against convection, and indeed, small transient convective cells do form within the upper envelope, as also seen in e.g., \cite{um16} and \cite{tyr17}. Accretion then nearly halts before three later dramatic bursts that are separated by long quiescent phases. During each pause, the star thermally relaxes, and convective cells formed in the outer envelope in each accretion burst eventually merge with the central region as the star becomes almost wholly convective.

The last of these bursts corresponds to the collision of the main accretion disk with a smaller, satellite disk that likely hosts another SMS.  Our {\sc enzo} simulations lack the resolution to determine if these two stars merge, so we simply include the collision of the smaller disk in the accretion rate of the main disk in our {\sc Kepler} runs. In this halo, we find that the star reaches $\sim$ 170 k\Ms\ at $\sim$ 1.2 Myr, at which point it collapses due to the GR instability while still burning hydrogen, with a central hydrogen fraction of $X_{\rm{c}} \approx 0.29$. We stress, however, that this merger event should be examined in greater detail, using e.g., a detailed three-dimensional smoothed particle hydrodynamical simulation to follow the trajectories of the stars and their subsequent interaction. We will study such systems in future work. 
Notably, however, we can confirm that the two massive stars in \texttt{Halo 01} will interact while both are on the main sequence, ultimately colliding in a supermassive stellar merger or forming a supermassive binary.

The SMS in \texttt{Halo 02} initially only accretes at an average rate of about 0.07 \Ms\ yr$^{-1}$ for the first 500 kyr. This leads to a (relatively) shallow radiative envelope \citep[in keeping with previous calculations, e.g.,][]{tyr17, hle18b}. Then the disk fragments, and accretion thereafter proceeds in an extremely clumpy manner, with the first accretion spike at $\sim$ 500 kyr leading to the emergence of a deep high-entropy radiative envelope, which then largely relaxes over the subsequent, relatively quiescent $\sim$ 0.3 Myr before the next accretion spike.  During this spike, the supermassive star rapidly accretes 40 k\Ms\ over $\sim$ 100 kyr in a marginally-convective, superadiabatic envelope before accretion nearly halts. Over the next 200 kyr the star largely thermally relaxes again, notably with a strong gradient between the inner convective core and the outermost envelope, before collapsing while still burning hydrogen ($X_{\rm{c}}$$\approx 0.34$) 
at the onset of a third accretion burst at 1.46 Myr at a final mass of $\sim$ 155 k\Ms. This is extremely close to the upper boundary mass for avoiding direct collapse for fully thermally relaxed SMSs \citep{tyr20a}.

In \texttt{Halo 08}, after a brief initial burst that peaks at 0.37 \Ms\ yr$^{-1}$, accretion proceeds at rates of 0.1 - 0.3 \Ms\ yr$^{-1}$ for the first $\sim$ 700  kyr of the star's evolution. During this time, the SMS very closely resembles stars accreting at uniform rates with similar average values \citep[e.g.,][]{hos13, tyr17, hle18b}. This is only somewhat interrupted by a minor accretion episode at $\sim$ 0.7 - 0.8 Myr, which peaks at just below 0.5 \Ms\ yr$^{-1}$ before subsiding. Although convective cells form in the outer envelope, the supermassive star remains far from being thermally relaxed up to its collapse at 0.954 Myr and $\sim$ 186 k\Ms, when it is still burning hydrogen ($\rm{X}_{\rm{c}} \approx0.38$).

The SMS in Halo~10 grows by nearly 0.4 \Ms\ yr$^{-1}$ for the first 100 kyr of evolution, with the accreted mass mostly remaining in the high-entropy envelope. Accretion then begins to slow, allowing the star to thermally relax somewhat and the central, nuclear-burning convective region to grow both in absolute mass and in total mass fraction of the star.  The star continues to grow at a more modest pace over the next 1.6 Myr, which allows its convective hydrogen- and helium-burning ``core'' mass fraction to grow as it continues to thermally relax. The mean accretion rate over the life of the SMS is $\sim$ 0.07 \Ms\ yr$^{-1}$. During this latter accretion phase, transient convective regions within the high-entropy envelope arise and disappear or merge with the central convective region before accretion nearly halts at 1.7 Myr.  A little over 100 kyr later, the star is almost fully convective. Shortly after the onset of a final accretion burst at $\sim 1.9$ Myr, the SMS collapses near the end of central hydrogen-burning ($\rm{X}_{\rm{c}}\approx0.06$) at a final mass of 132 k\Ms.

As in halos 01 and 02, the disk in \texttt{Halo 12} is strongly turbulent and begins to fragment soon after formation, which leads to large accretion bursts at the beginning of our {\sc Kepler} run and at 0.5 Myr and 0.65 Myr.  Unlike those in the other halos, the SMS in \textbf{Halo 12} produces somewhat more substantial, transient semi-convective regions at the interface between the convective core and the high-entropy envelope, the reason for which remains unclear. Ultimately, the precise treatment of convection in extremely radiation-dominated SMSs remains highly uncertain, with potential repercussions for the stellar structure, the response to variable accretion, and the dredge up and mixing of metal-enriched matter in the stellar envelope; however, a full exploration of 3-D convective motions in SMS envelopes is beyond the scope of the present work.  We find that the star collapses at 1.22 Myr at a final mass of 178 k\Ms\ while burning hydrogen ($\rm{X}_{\rm{c}} \approx 0.31$).  At collapse, the star has not been able to fully thermally relax because of persistent rapid accretion and is therefore not quite fully convective.

The accretion history in \texttt{Halo 16} is somewhat of an outlier in our sample because the initial peak associated with the formation of the disk is particularly strong and persistent, depositing nearly 80 k\Ms\ onto the star in its first 200 kyr. Accretion then almost entirely halts before resuming at $\sim$ 700 kyr. Infall rates at the center of the disk fall to low values because \texttt{Halo 16} collides with three other halos just before cooling and collapsing. These mergers spin gas up at the center of the halo and make the disk very rotationally supported, reducing accretion onto the star. The initial accretion peak produces an enormous, high-entropy envelope surrounding the convective hydrogen-burning region of the star. The latter grows in mass much more slowly, following the thermal relaxation of the SMS, only nearly catching up with the total mass of the star at $\sim$ 400 kyr, well after the initial accretion burst has ended. After accretion resumes at 0.7 Myr, the SMS slowly accumulates another 30 k\Ms\ before collapsing at a mass of 110 k\Ms\ at 1.68 Myr. This is the only star in our sample that reaches core hydrogen exhaustion ($\rm{X}_{\rm{c}}\approx5\times10^{-12}$, with a central helium fraction of $\approx$ 100\%). 
Coincidentally, this is also roughly when a second disk forms in \texttt{Halo 16}, raising the possibility of interactions between a second SMS and the DCBH of the first star.  Such interactions could result in a tidal disruption event that could be highly luminous in the NIR today \citep{ki16} or a long-lived, interacting supermassive X-ray binary. We will investigate the evolution of the companion disk in \texttt{Halo 16}, and the possibilities for interaction with the DCBH, in future studies.

\begin{table*}
\begin{center}
\hskip-2cm
    \begin{tabular}{ccccccccc}
        Halo & $z_{col}$ & $M_{\rm{halo}}$ & spin & N & $M_{\rm{SMS}}$ & $\rm{X}_{\rm{c}}$ & $t_{\rm{GR}}$ & $\dot M_{\rm{avg}}$\\
         & & $10^{7}$ \Ms\ & & & $10^{5}$ \Ms\ & & Myr & \Ms\ yr$^{-1}$\\
         \hline
1 & 16.7 & 3.68 & 0.0389 & 0 & 1.73 & 0.29 & 1.17 & 0.15\\
2 & 14.5 & 8.47 & 0.0388 & 0 & 1.55 & 0.34 &1.46 & 0.11\\
8 & 20.4 & 1.15 & 0.0321 & 0 & 1.86 & 0.38 & 0.95 & 0.19\\
10 & 17.3 & 2.60 & 0.0500 & 1 & 1.32 & 0.06 & 1.95 & 0.07\\
12 & 16.8 & 1.93 & 0.0471 & 0 & 1.78 & 0.31 & 1.22 & 0.15\\
16 & 16.5 & 2.91 & 0.0258 & 3 & 1.10 & 0.00 & 1.68 & 0.06\\
19 & 13.9 & 2.12 & 0.0072 & 0 & 1.77 & 0.19 & 1.79 & 0.10\\
20 & 17.7 & 3.56 & 0.0199 & 0 & 1.78 & 0.14 & 1.48 & 0.12\\
    \end{tabular}
    \caption{Properties of the halos at collapse  (redshifts, masses, spin parameters, and number N of major mergers prior to collapse) and their stars (final masses and central hydrogen fractions, times at which they reach the post-Newtonian instability, and average accretion rates).}
    \label{tab:properties}
    \end{center}
\end{table*}

Accretion in \texttt{Halo 19} is similar to that in \texttt{Halo 01}: after the initial peak, accretion falls to modest rates with occasional and brief bursts prior to the appearance of a large burst due to the merger of the main disk with a satellite disk, in this case at 1.4 Myr.  Here again, although the satellite disk may also host an SMS we make no attempt to model its mass or follow its merger with the star in the main disk.  Because of the lower infall rates up to 1.4 Myr the SMS in the main disk has mostly thermally relaxed and become almost fully convective by the time the disks collide, and it remains largely convective during the collision. Given its relatively modest prior accretion rate, the SMS within the most massive disk in \texttt{Halo 19} is largely thermally relaxed and almost fully convective by the time of the merger, remaining largely convective as it survives the resulting burst of accretion.  After another much smaller spike in accretion at $\sim$ 1.6 Myr, the star finally collapses at 1.79 Myr at a mass of 177 k\Ms\ while still burning hydrogen ($\rm{X}_{\rm{c}}\approx0.19$).

The formation of the central disk in \texttt{Halo 20} is accompanied by a particularly strong accretion peak, similar to that in \texttt{Halo 16}. In \texttt{Halo 20}, however, accretion never subsides, in spite of the formation of another disk that soon rivals the original in mass. A close encounter with this companion disk at 1.36 Myr produces a second powerful surge in accretion \citep[see][for more details]{pat20a}. After this burst, accretion falls off and the SMS collapses at 1.48 Myr at a final mass of 178 k\Ms\ while still burning hydrogen ($\rm{X}_{\rm{c}}\approx0.14$).

\subsection{General Properties}

Comparing the results in Figure \ref{Kippenhahn} with the classic, constant accretion rate case examined previously \citep[see e.g.,][]{hos12, hos13, tyr17, hle18b}, it is immediately apparent that a major consequence of the significant variations seen in more realistic accretion histories is a much wider diversity of stellar structures throughout the evolution of SMSs. In some cases, we find SMSs almost wholly thermally relaxed, in particularly sharp contrast to the now-standard picture of a deep, radiative, high-entropy envelope surrounding a convective core \citep{begel10}. In principle, this invites the possibility that the SMS may become a significant source of photoionizing emission, and thus provide a significant source of feedback in the halo which may impede its further growth \citep{sak15, sak16b}. If so, this would indicate that the evolution of atomic cooling halos and SMSs is strongly coupled, and a careful consideration of the accretion history must be made in evaluating the reliability of any such halo simulation \citep[see also][]{latif2020d}.

Two limitations prevent us from making a more definite statement in this regard: first, radiative feedback is not included in the original calculations of \cite{pat20a}; second, the photospheric temperatures of accreting SMSs are generally found to be systematically hotter in Kepler simulations than found in other studies \citep[e.g.,][]{hos13, hle18b}, for reasons which remain uncertain but which may relate to problems with resolution near the surface. For a detailed discussion of benchmarking efforts, see \citealt{titans}. We note that the effective temperatures of SMSs modelled here never exceed $\sim 2$--$5\times 10^{4}$ K during their lifetimes, such that only the Wien tail of their emission would provide significant ionizing photons; this together with the particularly dense gas flows seen in the halos we consider suggests feedback may remain ineffective in these particular cases \citep[see also][]{sak20}, however we do not speculate further for the reasons outlined above. 

In many other aspects, however, these objects do resemble the standard evolution of SMSs.  We find that the object at the center of the disk in each halo survives initial thermal contraction and reaches pp-chain hydrogen-burning without first collapsing to a BH.
Eventually, the central temperature exceeds $\rm{T}_{\rm{c}}\sim 10^{8}$ K, permitting the onset of of the triple-alpha reaction, quickly producing the CNO abundances needed to reach sufficient hydrogen-burning to stabilizing the star against further contraction, and producing a long-lived SMS.  The subsequent evolution of the star depends on the halo's accretion and merger history but a number of general trends are evident.

All of the stars in our sample undergo collapse at some point during the hydrogen-burning main sequence with the nominal exception of the SMS in \texttt{Halo 16}, which collapses at the exhaustion of its central hydrogen.  This is in keeping with the large masses deposited onto these SMSs by rapid accretion, which in almost all cases bring the SMS up to the post-Newtonian limit and the onset of the GR instability before the end of their main sequence lifetimes \citep[as also found in, e.g.,][]{tyr17,hle18a}. 
The SMS in \texttt{Halo 16} accretes almost all its mass in its first 200,000 yr and then lives for very nearly the main sequence lifetime predicted for an Eddington-limited star, before collapsing at the age expected for a wholly thermally-relaxed SMS \citep{tyr20a}.  Note that some stars, like those in Halos 10 and 19, live somewhat longer than the Eddington-limited main sequence lifetime because of the ingestion of additional hydrogen at later times in their accretion histories.

\begin{figure*}
    \centering
    \includegraphics[width=0.95\textwidth]{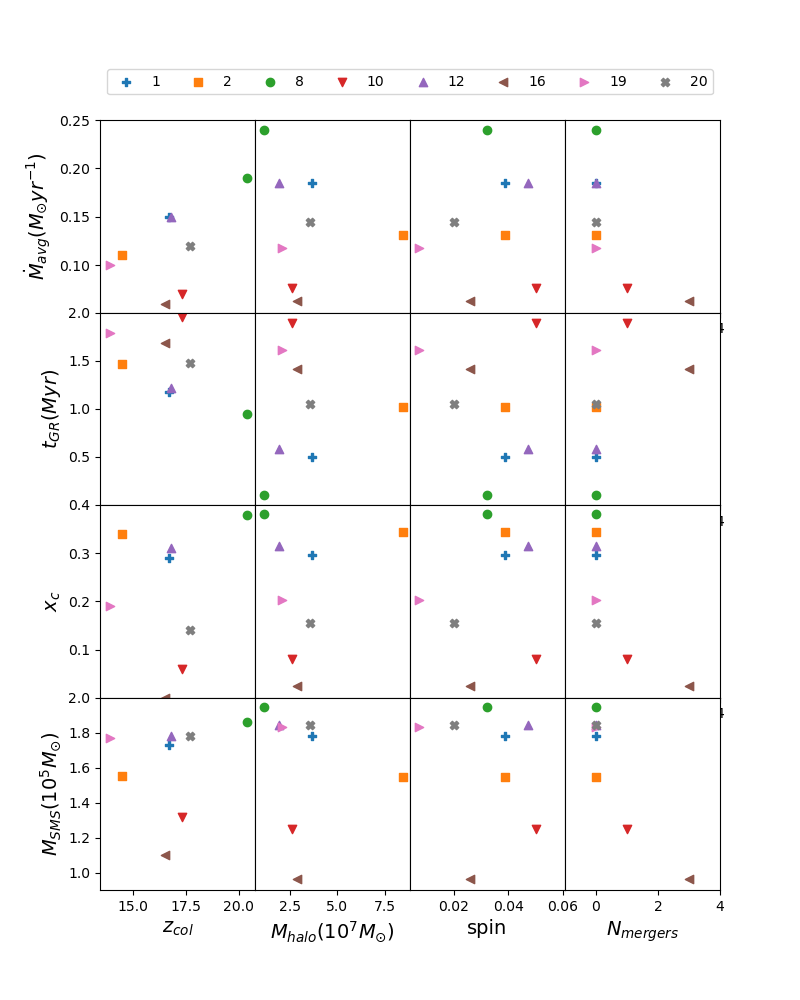}
    \caption{Scatter plots of mean accretion rates and final SMS masses, core hydrogen fractions and ages from our {\sc Kepler} models vs. key properties of the host halos from \citet{pat20a}. Individual halos are each identified by a unique colour and shape for all plots, as described in the figure legend. 
    }
    \label{fig:scatter}
\end{figure*}

The properties of each SMS and its host halo are listed in Table \ref{tab:properties}. Note that these  eight halos were chosen by \cite{pat20a} to span their likely range of spin parameters and assembly histories, i.e., growth primarily by accretion, by major mergers, or both.  Here, a major merger is defined to be the collision of two halos with mass ratios of 1/5 or more. 
We show scatter plots of mean accretion rate and final stellar mass, core hydrogen fraction and age versus halo mass, collapse redshift, spin parameter and number of major mergers prior to collapse in Figure~\ref{fig:scatter}.  Although our sample is small, it is clear that some properties of SMS populations at high redshift can be inferred from those of the halos in which they form.

First, a history of major mergers limits the maximum mass of the SMS and therefore produces the least massive DCBHs.  These halos form the smallest and most rotationally-supported disks, and thus have the lowest accretion rates.  Lower rates create more thermally-relaxed, compact SMSs with less opportunity for growth before significant nuclear evolution, as in \texttt{Halo 16} where the star collapses after reaching the end of the main sequence.  Second, if we exclude the two halos with major mergers prior to collapse, we see that there is also a tentative correlation between redshift and accretion rate, and thus an anti-correlation between redshift and SMS lifetime, because more rapidly-accreting SMS have shorter lives \citep{tyr17}. In the six halos that grew primarily by accretion, we also see a tentative anti-correlation between halo mass and final SMS mass, suggesting that more massive halos may produce less massive quasar seeds, however this is primarily motivated by the extreme case of \texttt{Halo 2}, which has about 3 times the mass of the other halos in our sample. Therefore, we defer further discussion of this trend to future studies with larger samples, as we cannot presently exclude some stochastic variation due to a random outlier. Perhaps the most striking feature with regard to the final mass, however, is the only modest deviations found given even highly variable accretion. In particular, taking the mean accretion rate over the lifetime of the star in each halo, and interpolating the final mass expected from previous constant accretion rate models \citep{tyr17}, we find the final mass in our constant accretion case differs stochastically, but only within a margin of $\sim 10\%$.

In some respects, \texttt{Halo 08} is the most extreme object in our sample because it collapses at the highest redshift and produces the most massive but shortest-lived star, although it is only slightly more massive than those in \texttt{Halo 12}, \texttt{Halo 19}, and \texttt{Halo 20}.  This means it is also the least-evolved at the time it encounters the GR instability and collapses with a central hydrogen fraction of 0.38. While we do not follow the ensuing collapse in detail because of the limitations of our 1D post-Newtonian approximation, if even a small fraction of nuclear-processed material were to be expelled to the surrounding environment it could provide a profoundly important source of chemical enrichment in the very early Universe, and an unusual one given the hot, hydrogen-rich burning conditions and the enormous mass available. Indeed, such a mechanism is not unlike that suggested to explain the unusual abundances of proton-rich elements observed in globular clusters \citep{Denissenkov14}. While in the present study we neglect rotation \citep[though see][]{hle18a}, this together with the existing uncertainties in mass loss and the nature of convection in supermassive envelopes, may yet allow for chemical enrichment by SMSs of their surrounding environment, however this is beyond the present scope.

Finally, we note that all but one of our stars (the one in \texttt{Halo 01}) collapse before the end of our {\sc enzo} simulations, well before the gas in the disk is exhausted. Accretion onto the nascent BH from its natal disk or a companion star will release X-rays that will alter the course of the evolution of the halo by, for example, catalyzing $\rm{H}_{2}$ formation and triggering new star formation in the vicinity of the BH \citep[e.g.,][]{aycin14,IT15}. 
Stellar evolution calculations with accretion rates from cosmological simulations will be required to predict when X-rays are turned on in a halo after the death of its SMS in future studies of its later evolution. 

\section{Conclusion}

Stars evolving in cosmological flows taken from numerical simulations of the collapse of atomically-cooling halos become supermassive before collapsing to BHs after 1 - 1.5 Myr. Accounting for the nuclear evolution and the short- and long-term response of SMSs to accretion, we find that all objects within our sample collapse during or at the end of the hydrogen-burning main sequence. In particular,  halos that cool purely by atomic cooling produce DCBHs with a narrow range of masses ($\sim$ 1 - 2 $\times$ 10$^5$ \Ms) for a wide range of collapse redshifts, spin parameters and merger histories. These masses are similar to the final masses of stars in \citet{tyr17}, whose uniform accretion rates were comparable to our average ones. Future studies incorporating radiative feedback and careful estimates of the SMS's ionizing photon luminosity will be able to investigate whether extended quiescent accretion phases within the disks formed in some halos may be able to modify this picture in some cases.

\citet{pat20a} do not bridge all the spatial scales of accretion onto the star because they are limited to resolutions of $\sim$ 0.01 pc, so it is possible that fragmentation
and less massive stars could appear at smaller scales (e.g., \citealt{bec15,bec18}).  Our SMS masses should therefore be taken to be upper limits.  In numerical simulations with higher resolutions, however, small clumps tend to merge over time at these scales.  Whether or not one star or many form in purely atomically-cooled halos remains an open question. 

Previous studies have shown that Pop III SMSs with masses similar to those in our models would  have extremely large luminosities that could be detected in the near infrared (NIR) by the {\em James Webb Space Telescope} ({\em JWST}) and large ground-based telescopes in the coming decade \citep{hos13,sur18a,sur19a,wet20a}.  Their BHs could also be found in the NIR at $z \sim 15 - 20$ using {\em JWST} \citep[with $\sim$24--100 hr exposures by NIRCam, as shown by][see therein for details]{pac15,nat17,bar18,wet20b} and at $z \sim$ 6 - 8 by {\em Euclid} and the {\em Roman Space Telescope} \citep[although lensing by galaxy clusters and massive galaxies in their wide fields could extend these detections up to $z \lesssim$ 10 - 15;][]{vik21a}.  DCBHs will only be marginally visible to the Square Kilometre Array or next-generation Very Large Array in the radio at $z \gtrsim$ 6 - 8 \citep{wet20a} but would become more luminous after growing to larger masses at later times.  They could also be detected out to $z \sim$ 10 by future X-ray missions such as the {\em Advanced Telescope for High-Energy Astrophysics} ({\em ATHENA}) and {\em Lynx} \citep{athena,lynx}.

Here, we have focused on the first disk to form in each halo, but \citet{pat20a} found that other disks can appear and exchange mass or even merge with the original disk. Although we do not model such mergers here, we find that they can occur at any time from when the SMS in the main disk is early in its evolution to well after it has collapsed to a black hole. This suggests a number of possible interactions between SMSs and their progeny in the early Universe, from supermassive stellar mergers to stable mass exchange in supermassive X-ray binaries (``SMBXs'') to DCBH--DCBH mergers, the latter being detectable by LISA out to redshifts of $\sim$15--20 \citep{Hartwig18}. Such massive and early DCBH mergers would be easily discriminated from the normal build-up of massive BH mergers, which arise only at somewhat lower redshifts in the course of hierarchical structure formation \citep{Sesana07}. The growing evidence that multiple very massive and supermassive stars may have been common in atomically-cooled halos \citep[see also][]{Chon18,Regan18a,Regan18b,Regan20,pat21b} indicates that these exotic interactions occurred in the early Universe.

\acknowledgments

We thank the referee for their insightful and helpful comments which improved the clarity of the manuscript. T.E.W.\ acknowledges support from the National Research Council Canada's Plaskett Fellowship. S.P.\ was supported by STFC grant ST/N504245/1 and D.J.W.\ was supported by the Ida Pfeiffer Professorship at the Institute of Astrophysics at the University of Vienna and by STFC New Applicant Grant ST/P000509/1. Our Enzo simulations were performed on the Sciama HPC cluster at the Institute of Cosmology and Gravitation at the University of Portsmouth. A.H.\ has been supported, in part, by the Australian Research Council (ARC) Centre of Excellence (CoE) for Gravitational Wave Discovery (OzGrav), through project number CE170100004; by the ARC CoE for All Sky Astrophysics in 3 Dimensions (ASTRO 3D), through project number CE170100013; and by the National Science Foundation under Grant No. PHY-1430152 (JINA Center for the Evolution of the Elements, JINA-CEE).

\bibliographystyle{aasjournal}

\end{document}